\documentclass[pre,superscriptaddress,twocolumn]{revtex4-2}

\usepackage{enumerate}
\usepackage{amsfonts,amssymb,amsmath}
\usepackage[]{graphics,graphicx,epsfig}
\usepackage{amsthm}
\usepackage{amsmath,amssymb}
\usepackage{graphicx}
\usepackage{dcolumn}
\usepackage{natbib}
\usepackage{color}
\usepackage{multirow}
\usepackage{ulem}

\begin{document}


\title{Transition from order to chaos in reduced quantum dynamics}
\author{Waldemar K{\l}obus}
\affiliation{Institute of Theoretical Physics and Astrophysics, Faculty of Mathematics,
Physics and Informatics, University of Gda{\'n}sk, 80-308 Gda{\'n}sk, Poland}
\author{Pawe{\l}  Kurzy\'nski}
\affiliation{Institute of Spintronics and Quantum Information, Faculty of Physics, Adam Mickiewicz University, 61-614 Pozna\'n, Poland}
\author{Marek Ku\'s}
\affiliation{Center for Theoretical Physics, Polish Academy of Sciences
Al. Lotnik{\'o}w 32/46, 02-668 Warszawa, Poland} 
\author{Wies{\l}aw Laskowski}
\affiliation{Institute of Theoretical Physics and Astrophysics, Faculty of Mathematics,
Physics and Informatics, University of Gda{\'n}sk, 80-308 Gda{\'n}sk, Poland}
\affiliation{International Centre for Theory of Quantum Technologies, University of Gda{\'n}sk, 80-308 Gda{\'n}sk, Poland}
\author{Robert Przybycie\'n}
\affiliation{Center for Theoretical Physics, Polish Academy of Sciences
Al. Lotnik{\'o}w 32/46, 02-668 Warszawa, Poland} 
\author{Karol {\.Z}yczkowski}
\affiliation{Institute of Theoretical Physics, Jagiellonian University, Lojasiewicza 11, 30-348 Krak{\'o}w, Poland}
\affiliation{Center for Theoretical Physics, Polish Academy of Sciences
Al. Lotnik{\'o}w 32/46, 02-668 Warszawa, Poland} 
\date{\today}


\begin{abstract}
We study a damped kicked top dynamics of a large number of qubits ($N \rightarrow \infty$) and focus on an evolution of a reduced single-qubit subsystem.
Each subsystem is subjected to the amplitude damping channel 
controlled by the damping constant  $r\in [0,1]$, which plays the role of the single control parameter. 
In the parameter range for which the classical dynamics is chaotic,
while varying $r$ we find the universal period-doubling behavior 
characteristic to one-dimensional maps:
period-two  dynamics starts  at $r_1  \approx 0.3181$,
while the next bifurcation occurs at  $ r_2 \approx 0.5387$.
In parallel with period-four oscillations observed for 
$ r \leq r_3 \approx 0.5672$,
we identify a secondary bifurcation diagram 
around $r\approx 0.544$,
responsible for a small-scale chaotic dynamics inside the attractor.
The doubling of the principal bifurcation tree 
continues until $r \leq r_{\infty} \sim 0.578$, which marks the onset of the 
full scale chaos interrupted by the windows of the oscillatory dynamics
corresponding to the Sharkovsky order.
 Finally, for $r=1$ the model reduces to the standard undamped chaotic kicked top. 
\end{abstract}

\maketitle


\section{Introduction}

Studies on classical nonlinear systems became of a great significance
due to the numerous applications
to physics, chemistry, biology and engineering \cite{Stro18}.
One of the key achievements of these early investigations
consists in understanding of the route from regular to
 chaotic dynamics \cite{ASY96,Ott02}. 
Furthermore, a link between chaotic dynamics,
defined by exponential sensitivity to initial conditions,
and emergence of fractal structures was established \cite{PJS92}.
Discovery of the Feigenbaum universality of the
period doubling scenario in one-dimensional systems
led to new insights concerning the nonlinear dynamics \cite{Fe78,Fe80,LM94}.

A lot of attention was paid to investigate properties of quantum analogues
of classically regular and chaotic systems \cite{BT77},
as their investigations helped to reveal fine 
connections between classical and quantum mechanics \cite{Gu90}.
Although the standard unitary quantum evolution is linear,
so no exponential sensitivity to initial conditions can be detected by a state-vector overlap, there exist quantum phenomena which reflect presence of classical chaos.
The study of these properties, called {\sl quantum chaology} \cite{Be89}
significantly improved our understanding
of the classical limit of quantized chaotic systems,
as numerous {\sl signatures of quantum chaos} were identified \cite{Ha90,HGK19}
and explained with help of the theory of random matrices
\cite{Me04} and theory of periodic orbits \cite{Cvi}.

Several studies of classically chaotic dynamics
and the corresponding unitary quantum evolution,
which takes place in a finite dimensional Hilbert space,
where performed with a model of kicked top \cite{HKS87}.
It describes a spin undergoing constant precession around a fixed magnetic field 
subjected to a periodic sequence of nonlinear kicks. 
The corresponding quantum system is described by a unitary evolution operator,
of a fixed dimension, $d=2j+1$, where the quantum number $j$ is set 
by the squared angular momentum operator $J^2$ and eigenvalue $j(j+1)$.
If the kicking strength parameter $\beta$ is large enough
the classical dynamics on a sphere becomes chaotic
and the spectral properties of the unitary evolution operator $U$
can be described by an appropriate ensemble of random unitary matrices
\cite{KSH87}.
An apparent contradiction between exponential divergence
of neighboring trajectories of a chaotic classical dynamical system
and the linear evolution of the corresponding quantum system
can be explained by the fact that the limit time to infinity, necessary to define 
the Lyapunov exponent, and the limit $j\to \infty$,
corresponding to the classical limit of quantum theory, $\hbar \to 0$,
do not commute \cite{HGK19}. 

Investigations of quantized chaotic dynamics are relevant not only for
quantum theory but have also applications in several branches of
experimental physics \cite{St99}.
In particular, the model of quantum kicked top,
motivated by an experimental work of Waldner et al. \cite{WBY85},
 was later studied experimentally \cite{CSAGJ09,KABM19}.
The latter reference concerns nuclear magnetic resonance
experiments simulating the model of coupled kicked tops,
earlier analyzed in \cite{MS98,BL04,DDK04,TMD08}.

A physical realization of any model quantum dynamics
is subjected to dissipation and decoherence.
Although the original model of the quantum kicked top
is described by unitary time evolution \cite{HKS87},
it was later generalized \cite{Bra01,HGK19}
to take into account also effects of 
 dissipation and decoherence.

The model of quantum kicked tops were used to analyze 
properties of entanglement in coupled chaotic systems
\cite {FMT03,WGSH04,PLQ19,HKFB20}.
Although the dynamics of the entire bi-partite system is unitary,
the dynamics of the reduced state corresponding to a given subsystem
is non-unitary. 
Under assumption of a strong coupling between subsystems,
classically chaotic dynamics of individual tops,
and large dimension of the system, 
the partial traces of the composite system display
statistical properties characteristic of random density matrices \cite{PPZ16}.

As the Heisenberg 
time evolution of an isolated quantum state
is unitary and linear, $\rho \to U\rho U^{\dagger}$,
some non-linear effects may arise due to interaction with other
subsystems. For instance, the quadratic term, $\rho^2$ 
corresponds to quantum measurements performed on two copies of the
same quantum state $\rho$ \cite{BPTC09}.
Other models include nonlinear transformations,
in which individual entries of the density matrix are squared \cite{BPHG98}
and measurement based nonlinear rotation of the Bloch sphere \cite{HS01}.

In this work we are going to analyze a system of $N$ interacting qubits, 
described in the Hilbert space of a finite dimension $d=2^N$.
Dynamics of a single qubit represents kicked top in the chaotic regime,
(kicking strength $\beta = 6$),
and all the subsystems are coupled by an interaction Hamiltonian.
Therefore, the reduced dynamics of a 
qubit subsystem becomes effectively nonlinear as $N\rightarrow \infty$.

The aim of this contribution is to analyze properties of the non-linear dynamics
of a single qubit, obtained by partial trace over remaining subsystems,
under a realistic assumption that each subsystem is
subjected to the amplitude damping channel.
We demonstrate that depending on the value of the damping parameter $r$, equal for all subsystems, 
the dynamics of the reduced state exhibits various forms of very complex behaviors.
In particular, we show under what conditions the single qubit dynamics 
converges to a stable fixed point and when bifurcation occur.
Furthermore, we demonstrate that the period doubling scenario,
originally observed for classical systems \cite{Fe78,Fe80},
can be also applied to reduced dynamics of a damped quantum system.
In such a way the Feigenbaum route to chaos can be now 
identified also for quantum systems.
Apart of the standard period doubling scenario,
inside the period-two and period-four oscillatory dynamics,
we observe selfsimilar structure of higher order bifurcation diagrams,
responsible for a small-scale chaos inside the atractor.
Similar structures were observed for the classical, two-dimensional  
H{\'e}non map \cite{ZRSM00}.

Complementary goal of this project concerns investigation
of the purely quantum regime of the model obtained for a finite number of 
qubits. As fractal structures, characteristic to classical chaotic dynamics
become blurred by quantum effects \cite{WBBZ00,LZS03,Ja14},
it is particularly interesting to observe how the fine effects
related to classical period doubling scenario and strange attractors
get dominated by quantum effects.
Let us emphasize here that the model of damped coupled kicked tops,
investigated in this work,
can be related to physics of 
many body systems and interacting cold atoms.

This work is organized as follows. In Section~II we
introduce the model of damped coupled kicked tops
and present some of its properties. 
Fixed points of the system describing the dynamics of single qubit,
under the assumption of a large total number $N$ of qubits,
is presented in Section~III. In Section~IV we fix two parameters of the 
unitary evolution, 
so the system is solely described by the parameter $r$
governing the amplitude damping,
as $1-r$ can be interpreted as the damping strength.
The fixed parameters are chosen in such a way
that in the unitary limit,  $r \to 1$,
 the system becomes equivalent to the standard chaotic kicked top \cite{HKS87}.
Period doubling scenario for such a non-linear quantum system 
is analyzed in Section~V while strange attractors are investigated in Section~VI.
In Section~VII we study bifurcation diagrams and identify the windows of periodicity and in Section~VIII we study Lyapunov exponents. Concluding remarks are presented in Section~IX,
while the derivation of the effective single-qubit dynamics
in the limiting case $N \to \infty$ 
is provided in  Appendix.



\section{The model quantum system}

We consider a collection of $N$ interacting qubits. They are initially in a symmetric product state $\rho_0^{\otimes N}$ and we assume the following interaction Hamiltonian 
\begin{equation}
H = \frac{g}{2(N-1)} \left( \sum_{n=1}^N \sigma_z^{(n)}\right)^2, 
\end{equation}
where $\sigma_z^{(n)}$ is the Pauli-Z operator acting on the n-th qubit and $g$ determines the interaction strength. If $g={\mathcal O}(1)$ and $N\rightarrow \infty$, then each qubit from this collection undergoes an effective nonlinear unitary dynamics $U(\rho)\rho U^{\dagger}(\rho)$ (see Appendix A), where
\begin{equation}
U(\rho) = e^{-i\frac{\beta}{2}\langle \sigma_z \rangle\sigma_z},
\end{equation}
$\langle \sigma_z \rangle = \text{Tr}\{\rho \sigma_z\}$, $\beta=g\tau$ and $\tau$ is the time of interaction.

Next, we modify the evolution analyzed. The map is going to consist of three subsequent operations: (1) the above nonlinear unitary evolution $U(\rho)$,
 (2) local rotation of each qubit about y-axis, (3) amplitude damping to $|0\rangle$ state. The operations (1) and (2) generate the standard kicked top dynamics
\cite{HKS87,KSH87}  described by
\begin{equation}
V(\rho) = e^{-i\frac{\alpha}{2}\sigma_y}U(\rho),
\end{equation}
where $\alpha$ is the angle of rotation about y-axis. The total evolution is given by
\begin{equation}\label{dkt}
\rho_{t+1} = K_1V(\rho) \rho_t V^{\dagger}(\rho)K^{\dagger}_1 +  K_2 V(\rho) \rho_t V^{\dagger}(\rho)K^{\dagger}_2.
\end{equation}
In the above, 
\begin{equation}
K_1 = \begin{pmatrix} 1 & 0 \\ 0 & \sqrt{r} \end{pmatrix},~~~~K_2 = \begin{pmatrix} 0 & \sqrt{1-r} \\ 0 & 0 \end{pmatrix},
\end{equation}
are the amplitude damping Kraus operators \cite{NC2010},
which satisfy the desired  identity resolution,
$\sum_{i=1}^2 K_i^{\dagger} K_i={\mathbb I}$,
equivalent to the trace preserving condition.
The parameter $r\in [0,1]$ describes the degree of
damping in the model:  
for $r=1$ the operator $K_2$ vanishes,
so $r'=1-r$ playes the role of  the damping strength. 

After $t$ steps the state of the qubit is given by
\begin{equation}
\rho_t = \frac{1}{2}\left( \openone + x_t \sigma_x + y_t \sigma_y + z_t \sigma_z\right),
\end{equation}
where $\mathbf{v}_t = (x_t,y_t,z_t)$ is the corresponding Bloch vector. The evolution of $\mathbf{v}_t$ is determined by
\begin{eqnarray}
\label{xyz+}
x_{t+1} &=& \sqrt{r}[\left(x_t \cos(\beta z_t)-y_t \sin(\beta z_t)\right)\cos\alpha+ z_t \sin \alpha], \nonumber \\
y_{t+1} &=& \sqrt{r}[x_t \sin(\beta z_t) + y_t \cos(\beta z_t)], \\
z_{t+1} &=& 1+r[\left( y_t \sin(\beta z_t) -  x_t \cos(\beta z_t)\right)\sin\alpha + z_t \cos\alpha-1]. \nonumber 
\end{eqnarray}


\section{Fixed points and bifurcations}

Let $\mathbf{v}^{\ast} = (x^{\ast},y^{\ast},z^{\ast})$ denote a fixed point of the evolution (\ref{xyz+}). 
It follows
\begin{eqnarray}
x^{\ast} &=&\frac{\sqrt{r}\sin\alpha \left(1-\sqrt{r}\cos(\beta z^{\ast})\right)z^{\ast}}{1+r\cos\alpha - \sqrt{r} (\cos\alpha + 1)\cos(\beta z^{\ast})}, 
\nonumber  \\
y^{\ast} &=& \frac{r\sin\alpha\sin(\beta z^{\ast})z^{\ast}}{1+r\cos\alpha - \sqrt{r} (\cos\alpha + 1)\cos(\beta z^{\ast})},   \\
z^* &=& 1+rz^*\cos(\alpha)-r \nonumber \\
&+&r z^*\sin^2(\alpha) \frac{r -\sqrt{r}\cos(\beta z^*)}{1+r\cos(\alpha)-\sqrt{r}(\cos(\alpha)+1)\cos(\beta z^*)} \nonumber
\end{eqnarray}
This set of  equations is not easy to solve, so  we analyze them numerically. We define
\begin{eqnarray}
& & f(z^{\ast},r,\alpha,\beta) = -z^* + 1+rz^*\cos(\alpha)-r  \label{f}\\ 
&+&r z^*\sin^2(\alpha) \frac{r -\sqrt{r}\cos(\beta z^*)}{1+r\cos(\alpha)-\sqrt{r}(\cos(\alpha)+1)\cos(\beta z^*)}, \nonumber
\end{eqnarray}
and the goal is to look for solutions to $f(z^{\ast},r,\alpha,\beta) =0$.

The next goal is to investigate stability of these fixed points. To do this, we use the standard approach  \cite{Stro18}, i.e., we linearise the equations in a vicinity of a fixed point. More precisely, consider a small deviation from a fixed point
\begin{equation}
\mathbf{v}_{t}  = \mathbf{v}^{\ast} + \mathbf{\Delta v}_{t}.  
\end{equation}  
It follows
\begin{equation}
\mathbf{\Delta v}_{t+1} \approx \mathbf{A}_{\mathbf{v}^{\ast}}\mathbf{\Delta v}_{t},
\end{equation}
where $\mathbf{A}_{\mathbf{v}^{\ast}}$ is the Jacobian of the map at point $\mathbf{v}^{\ast}$. A fixed point $\mathbf{v}^{\ast}$ is stable if the modulus of all the eigenvalues of $\mathbf{A}_{\mathbf{v}^{\ast}}$ is not greater than one.


\section{Chaotic regime of the model}

From now on we fix the parameters of the model,
\begin{equation}\label{fix}
\alpha = \frac{\pi}{2},~~~~\beta = 6,
\end{equation}
as this choice leads to chaotic dynamics of the undamped kicked top 
in the classical limit  \cite{HKS87}. 
Therefore, the system is now described solely by the unitarity parameter $r$. Its behaviour in the two limiting cases is clear. For $r=0$ the system undergoes damping to $|0\rangle$ in one step, whereas for $r=1$ the is no damping in the system,
the evolution becomes unitary 
and reduces to the chaotic dynamics of the standard kicked top. 
These two extreme values correspond to two different behaviors -- order and chaos. Interesting things should happen in between and this is what we are going to examine below. 

\begin{figure}
	\includegraphics[width=8.5cm]{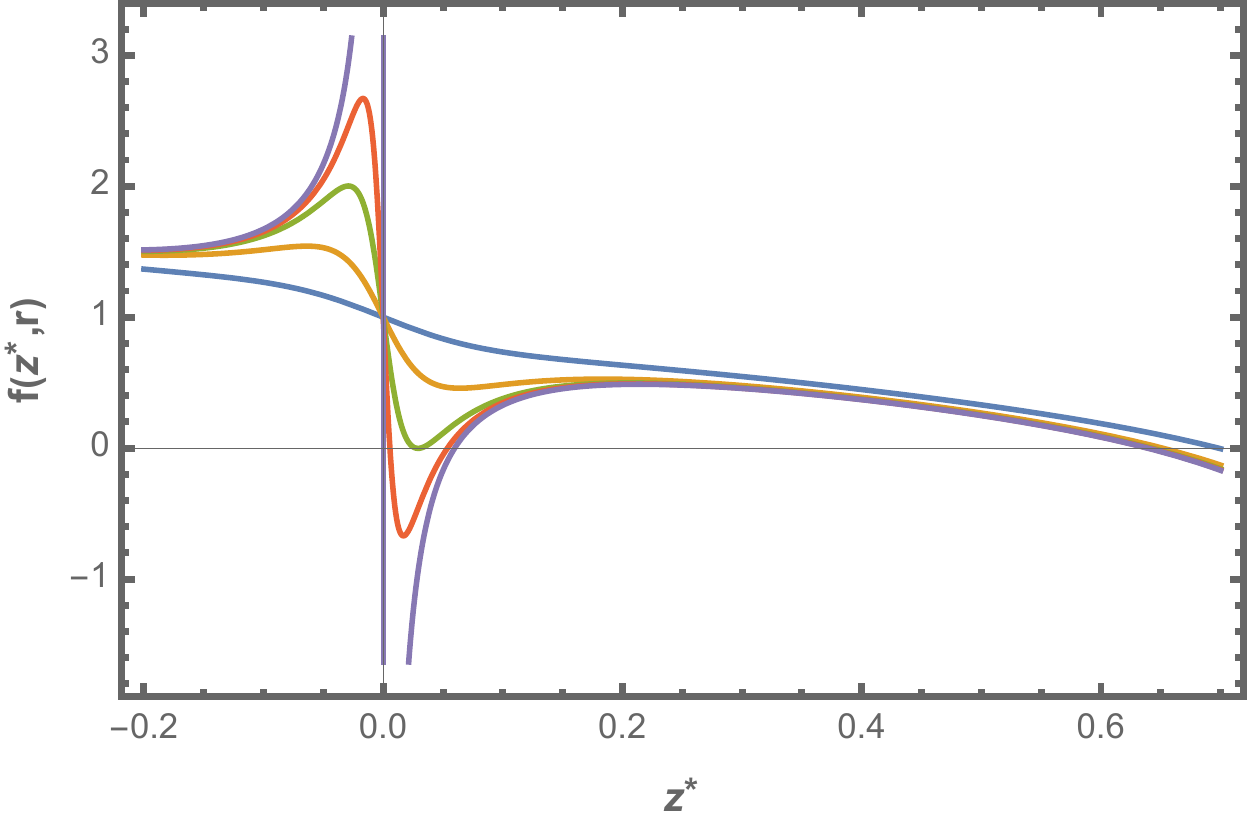}
\caption{The plot of $f(z^{\ast},r)$ for: $r=0.6$ (blue), $r=0.9$ (orange), $r=0.972\approx r_b$ (green), $r=0.99$ (red), $r=1$ (purple).} 
\label{f1}
\end{figure}

First, let us look for fixed points using $f(z^{\ast},r) =  f(z^{\ast},r,\pi/2,6)$ --
see Eq. (\ref{f}). 
We find that for $0 \leq r<r_{b}\approx 0.9719$ there is one fixed point,  denoted by 
$\mathbf{v}_{0}^{\ast}$. For $r_{b}<r<1$ there are three of them: $\mathbf{v}_{0}^{\ast}$, $\mathbf{v}_{1}^{\ast}$, and $\mathbf{v}_{2}^{\ast}$. The additional two appear in a saddle-node bifurcation. Finally, for $r=1$ there are two fixed points: $\mathbf{v}_{0}^{\ast}$ and $\mathbf{v}_{2}^{\ast}$. The fixed point $\mathbf{v}_{1}^{\ast}$ disappears due to discontinuity of $f(z^{\ast},1)$ at $z^{\ast}=0$ -- see Fig. \ref{f1}. 

Through the analysis of the corresponding Jacobian 
\begin{equation}
\mathbf{A}_{\mathbf{v}^{\ast}} = \begin{pmatrix} 
0 & 0 & -\sqrt{r} \\ 
-\sqrt{r}s(z^{\ast}) & \sqrt{r}c(z^{\ast}) & -\beta \sqrt{r}\left( y^{\ast} s(z^{\ast}) + x^{\ast} c(z^{\ast})  \right) \\
rc(z^{\ast})  & rs(z^{\ast}) & \beta r\left( y^{\ast} c(z^{\ast}) - x^{\ast} s(z^{\ast}) \right) \\
\end{pmatrix},
\end{equation}
where
\begin{equation}
s(z^{\ast}) \equiv \sin( \beta z^{\ast}), ~~~~c(z^{\ast}) \equiv \cos( \beta z^{\ast}),
\end{equation}
with $\beta=6$, we find that for $r \leq r_1\approx 0.3181$ the single fixed point $\mathbf{v}_{0}^{\ast}$ is stable, whereas for $r > r_1$ it becomes unstable as a result of a flip bifurcation \cite{Stro18}. On the other hand, for $r>r_b$ the new fixed point $\mathbf{v}_1^{\ast}$ is stable and $\mathbf{v}_2^{\ast}$ is unstable. In addition, for $r=1$ the fixed point $\mathbf{v}_{0}^{\ast}$ is unstable and the stability of $\mathbf{v}_{2}^{\ast}$ cannot be determined due to the fact that all eigenvalues of $\mathbf{A}_{\mathbf{v}_2^{\ast}}$ are equal to one. However, since the value $r=1$ corresponds to the standard kicked top, we know that $\mathbf{v}_{2}^{\ast}$ cannot be stable. Finally, the value of $z_0^{\ast}$ equals one for $r=0$ and monotonically decreases to $\approx 0.639$ for $r=1$, whereas the values of $z_1^{\ast}$ and $z_2^{\ast}$ are close to zero ($z_1^{\ast}<z_2^{\ast}<0.06$).


\section{Period-doubling and universality}

For $r_1<r<r_b$ there are no stable fixed points. At $r=r_1$ we observe the onset of period-2 oscillations, i.e., after a transient stage the state-space of the system becomes limited to just two points and the evolution flips one point to the other. As $r$ increases, the period of oscillations doubles. Interestingly, for $r_{s_1} \approx 0.5378 < r < r_{s_2}  \approx 0.5455$ we observe a departure from the standard period-doubling behaviour and 
emergence of higher-order bifurcation trees, which lead
 to a weakly chaotic dynamics inside the attractor. This self-similar structure
is discussed in more details in Section  \ref{selfsim}.

\begin{figure}
	\includegraphics[width=4.25cm]{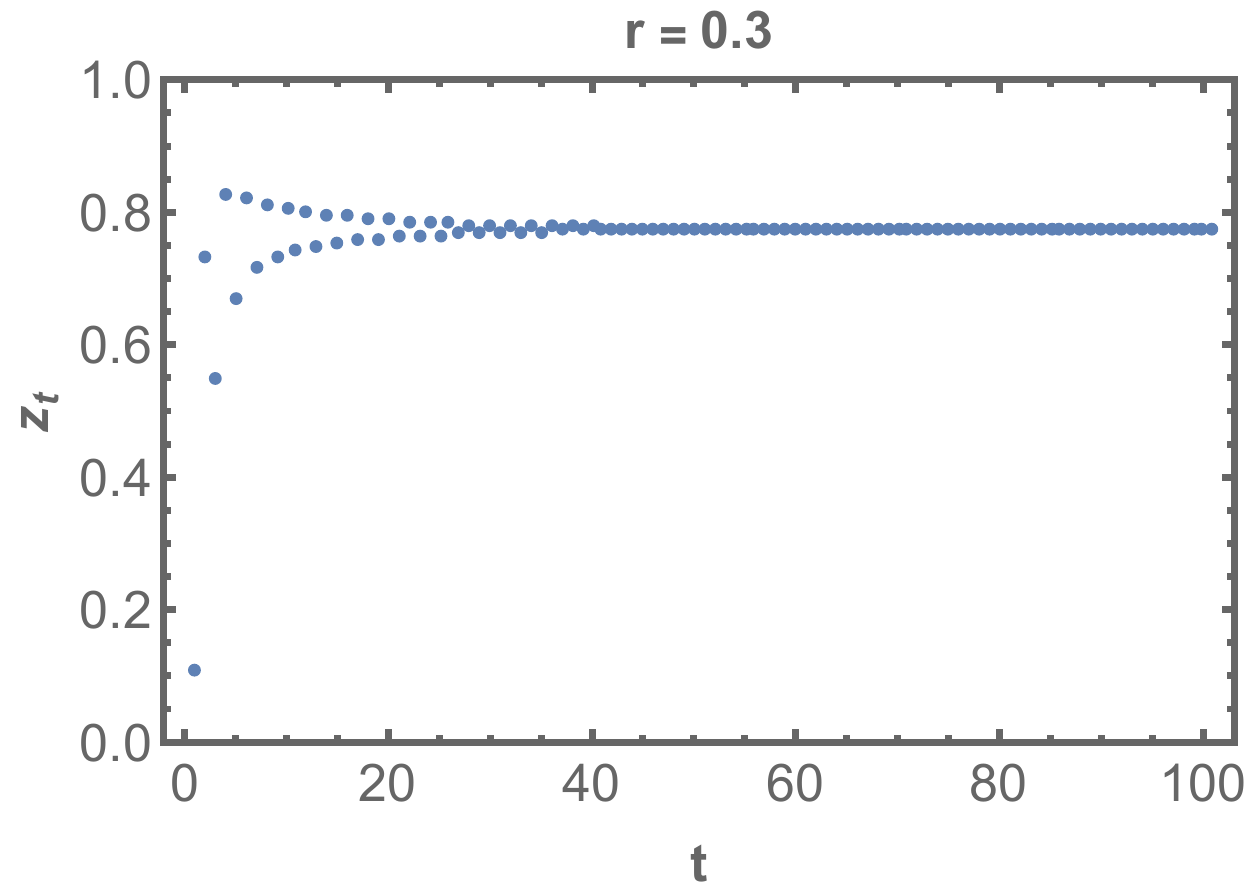} 
\includegraphics[width=4.25cm]{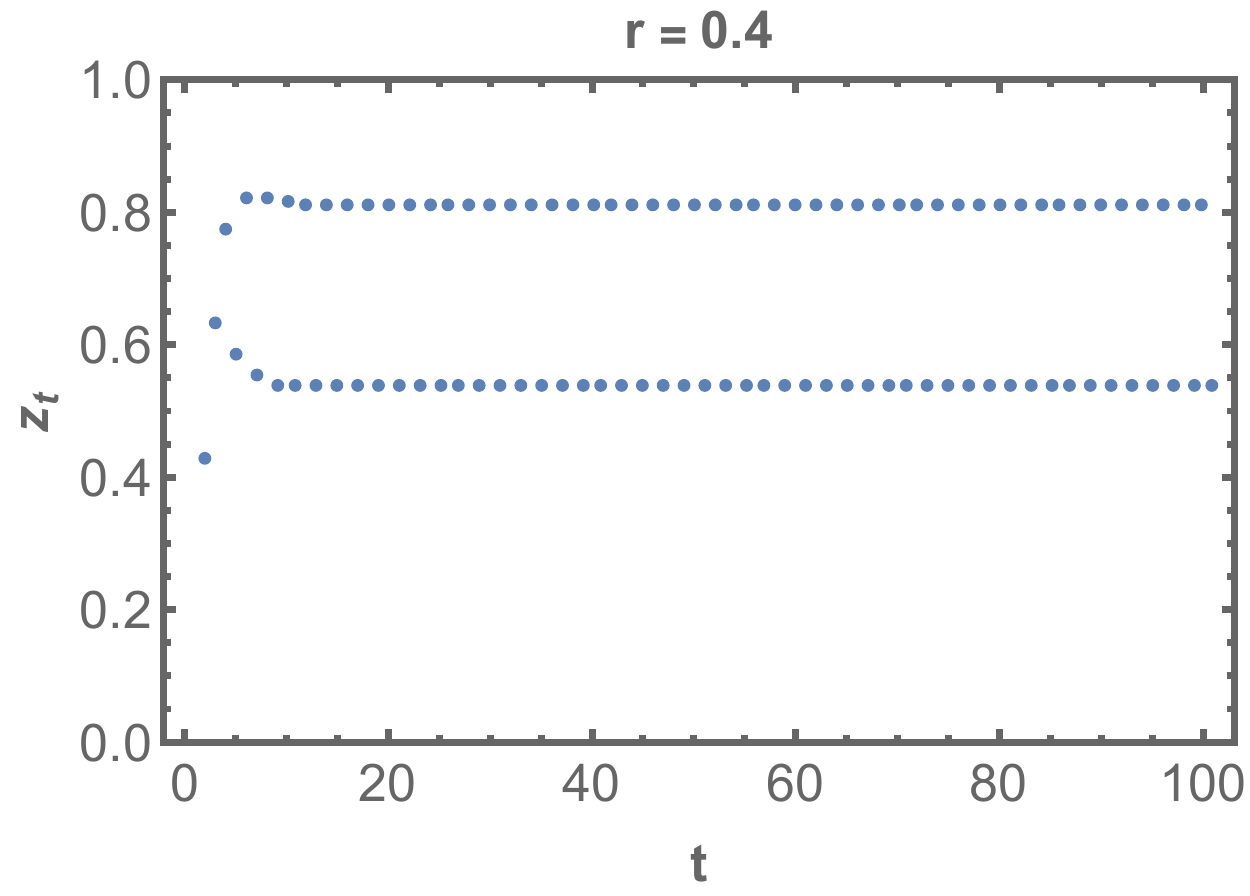}
\includegraphics[width=4.25cm]{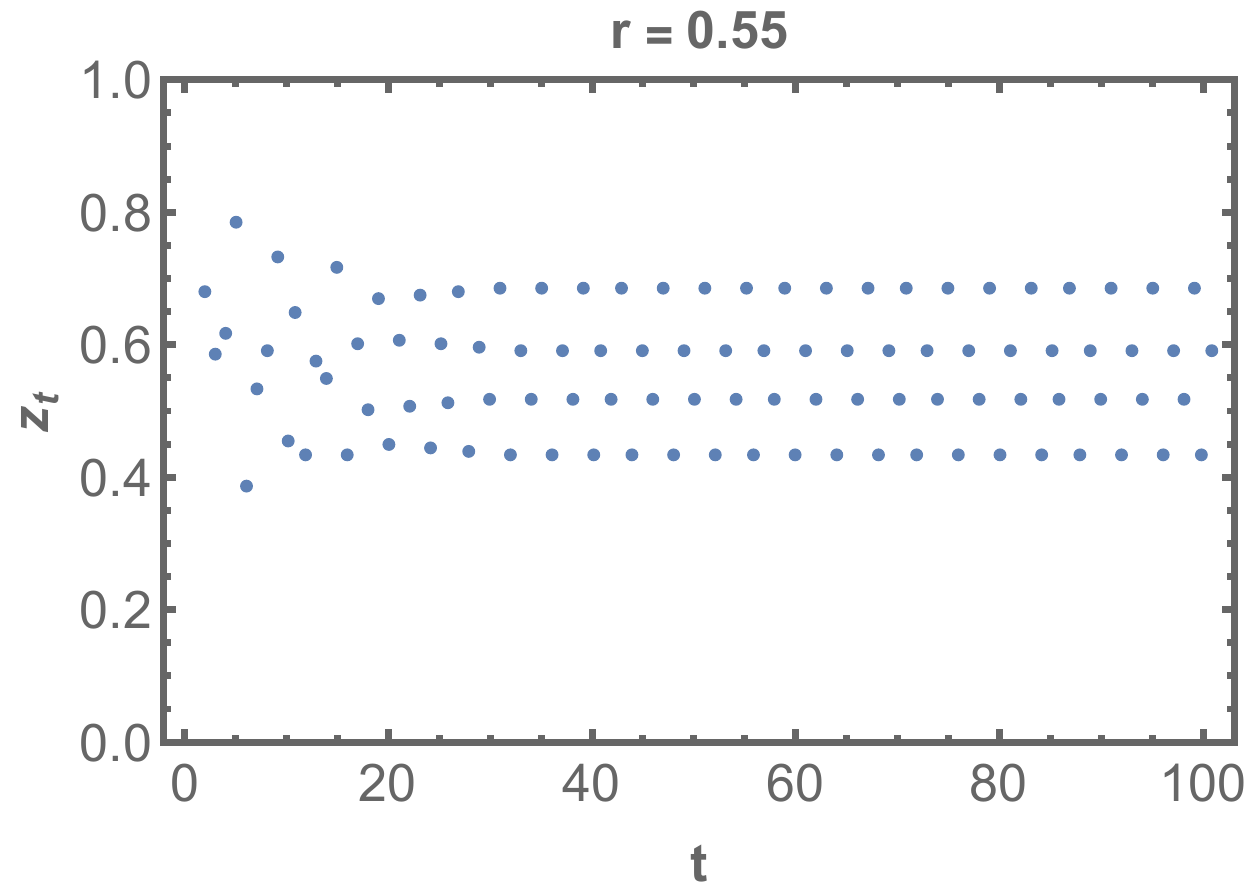} 
\includegraphics[width=4.25cm]{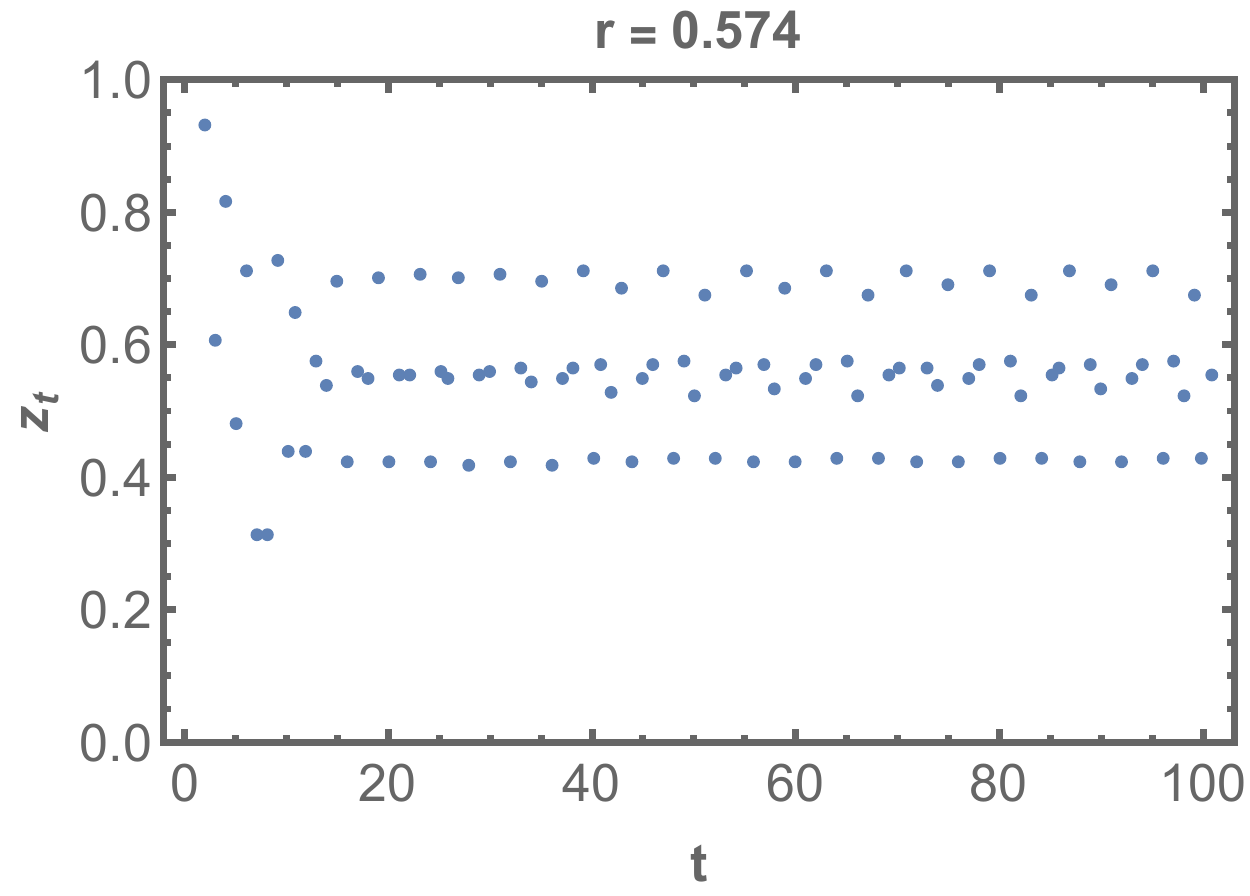}
	\includegraphics[width=4.25cm]{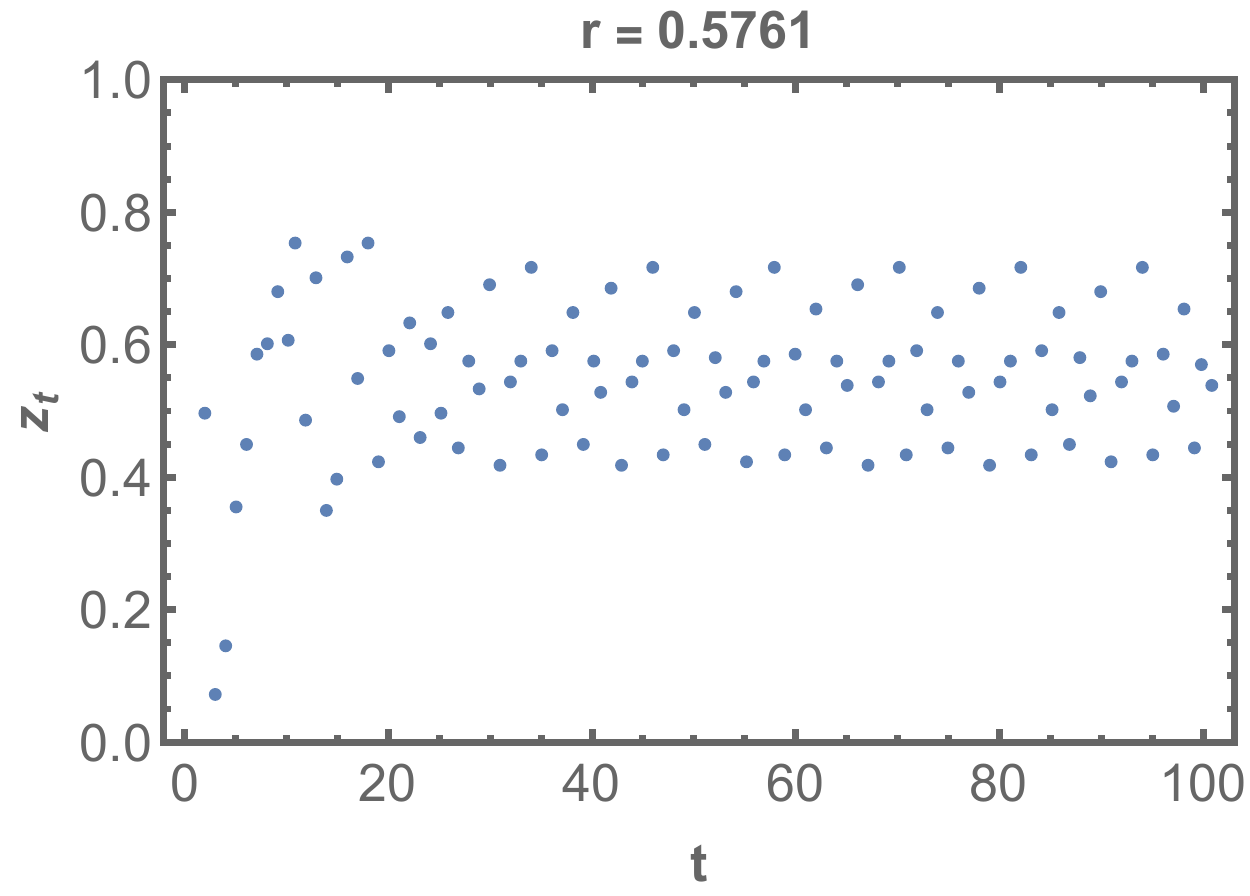} 
	\includegraphics[width=4.25cm]{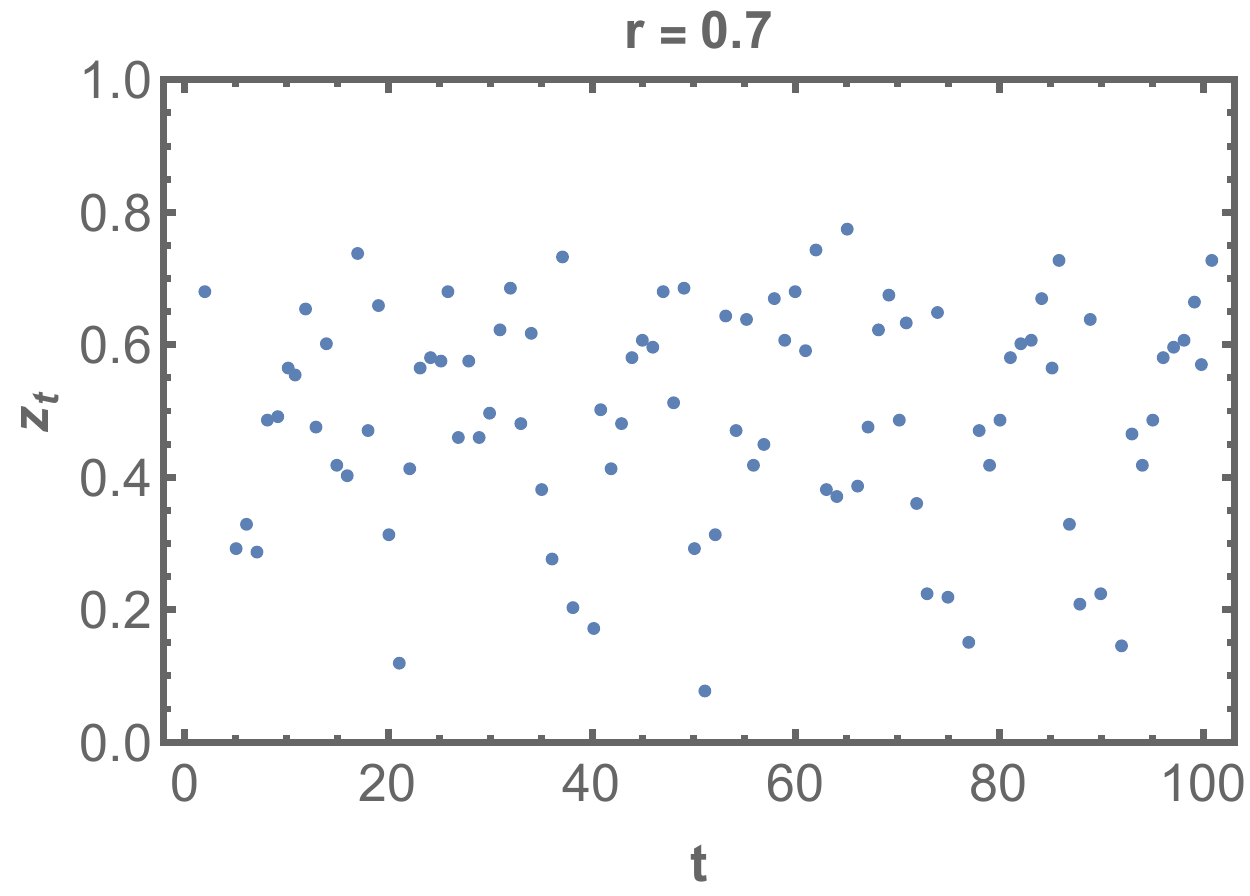}
\caption{The first 100 steps of the evolution of $z_t$ for different values of $r$ and a random initial state.} \label{f2}
\end{figure}

Examples of the evolution of $z_t$ for six different values of $r$ are presented in Fig. \ref{f2}. The dependence of the asymptotic behaviour on the parameter $r$ is summarised in Table \ref{t1}. The value of $r_{\infty}$, at which the onset of chaos occurs, is hard to determine in numerical experiments, since it is not easy to distinguish between multi-period oscillations and the irregular chaotic dynamics. However, we are going to upper bound it in a moment.

\begin{table}[t]
\begin{tabular}{|l|l|}
\hline
range of $r$ & behaviour \\
\hline
\hline
$0\leq r<r_1\approx 0.3181$ & stationary   \\
\hline
 $r_1<r<r_2\approx 0.5387$ & period-2   \\
 \hline
 $r_{s_1} \approx 0.5378 < r < r_{s_2}  \approx 0.5455$ & self-similarity \\
 \hline
 $r_2<r<r_3\approx 0.5672$ & period-4   \\
 \hline
 $r_3<r<r_4\approx 0.5729$ & period-8   \\
 \hline
 $r_4<r<r_5\approx 0.5741$ & period-16   \\
 \hline
 \ldots & \ldots \\
 \hline
 $r_{\infty} < r < r_b \approx 0.9719$ & chaos \\
 \hline
 $r_b<r<1$ & stationary   \\
\hline
 $r=1$ & chaos (kicked top)   \\
\hline
\end{tabular}
\caption{Asymptotic behaviour of the model for different values of $r$. \label{t1}}
\end{table}

The phenomenon of period-doubling was first observed in the logistic map 
 \cite{Fe78,Fe80}, but later it was found to occur in a large family of iterated maps that are described by a single parameter $r$ -- see \cite{LM94,Ott02}.
  For all these maps one can define a value $r_k$ that marks the onset of period-$2^k$ oscillations. Interestingly, Feigenbaum found a universal scaling behaviour, namely that the ratio $\frac{r_{n}-r_{n-1}}{r_{n+1}-r_n}$ tends to a constant value as $n$ goes to infinity
\begin{equation}
\lim_{n\rightarrow \infty} \frac{r_{n}-r_{n-1}}{r_{n+1}-r_n} = \delta = 4.669 201 609 \ldots
\end{equation}
The number $\delta$ is now known as the {\sl Feigenbaum constant}. 

The universal period-doubling behaviour and the approximate convergence to the Feigenbaum constant was observed in a number of one-dimensional physical systems and mathematical models \cite{Fe78,Fe80}. Below we demonstrate how an approximate convergence 
occurs in the three dimensional model analyzed here.
\begin{table}[h]
\begin{tabular}{|c|l|}
\hline
 ratio & value \\
\hline
\hline
$(r_2-r_1)/(r_3-r_2) $ & $\approx 7.74$   \\
\hline
$(r_3-r_2)/(r_4-r_3) $ & $\approx 5.0$ \\
 \hline
$(r_4-r_3)/(r_5-r_4) $ & $\approx 4.75$ \\
 \hline
\end{tabular}
\end{table}

The exact convergence to $\delta$ cannot be observed due to finite precision of numerical simulations. We estimated the values
of the parameters $r_k$ for $k=1\dots 5$ up to the order $10^{-4}$. 
 Assuming that 
\begin{equation}
\frac{r_{5}-r_{4}}{r_{6}-r_{5}} \approx \delta,  
\end{equation}
we can estimate $r_6 \approx 0.5743$, which allows us to conjecture that $r_{\infty} < 0.578$. This is confirmed in numerical simulations.

\begin{figure}
      \includegraphics[width=8.5cm]{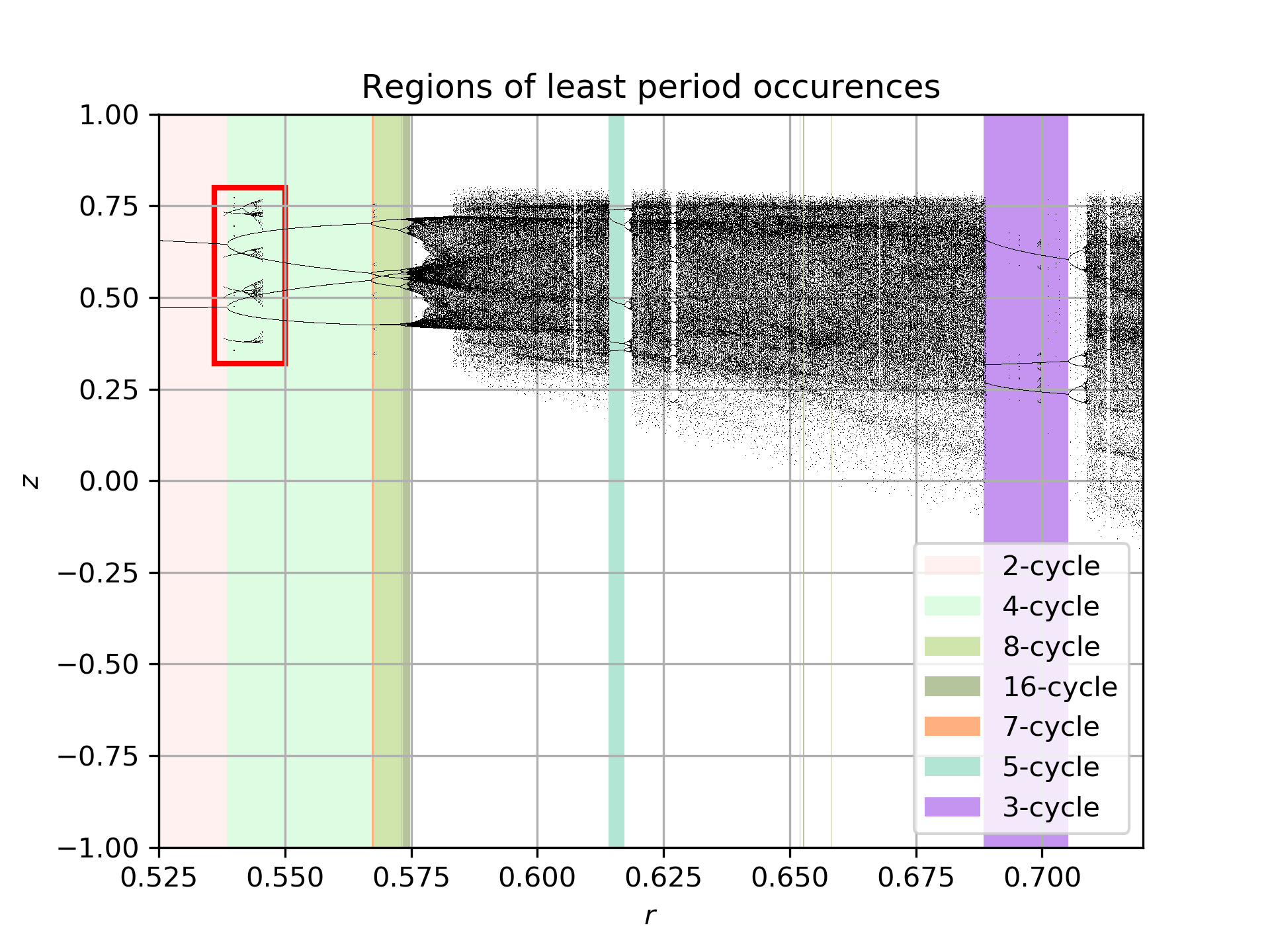}
	\caption{Bifurcation diagrams for the $Z$ coordinate of the Bloch vector
	  with cycles identified according to the Sharkovsky order.
	     Note secondary bifurcation diagrams located inside the 4-cycle around 
	  $r \approx 0.545 $ and magnified in Fig. \ref{self3}.
	 } 
	 \label{cykle2}
\end{figure}

  At a first glance the bifurcation diagram  presented in Fig. \ref{cykle2}
 is similar to the one of a logistics map 
   and of other systems that exhibit period-doubling behavior \cite{Fe78,ASY96,Ott02}. It also shows yet another universal property of such systems -- emergence of windows of periodicity, i.e., existence of regions in which the chaotic behaviour ceases and periodic behaviour re-emerges for some narrow regions of $r$. We find two transparent such windows in our system. The first one (narrow with 5 and 10-cycles) appears at the range $ 0.614 \leq r \leq 0.619$ and the second one (wider with 3 and 6-cycles) appears at the range $0.689 \leq r \leq 0.709$. This is in agreement with the celebrated Sharkovsky ordering,
$1 \prec  2 \prec 4 \prec 8  \prec  \dots \prec  7 \prec   5  \prec 3$,
see \cite{Sha64, LY75,Sha95}.

\section{Chaos and strange attractor}

\begin{figure}
\includegraphics[width=8cm]{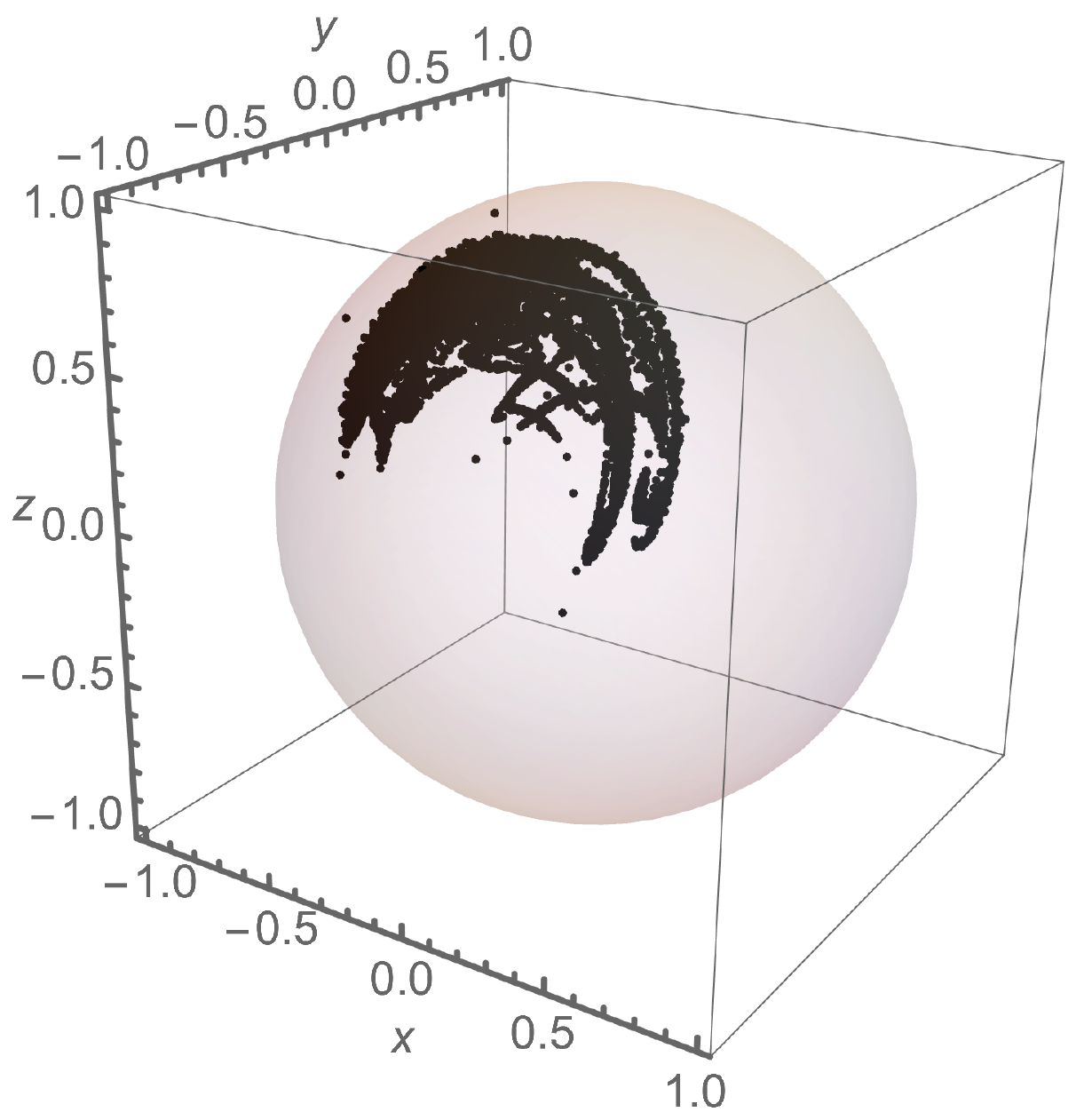}
\caption{Visualisation of a strange attractor:
 10000 steps of a trajectory
 stemming from a random initial state dynamics for $r=0.75$. 
 Black points denote subsequent positions of the Bloch vector.} 
\label{f3}
\end{figure} 

\begin{figure}
	\includegraphics[width=8cm]{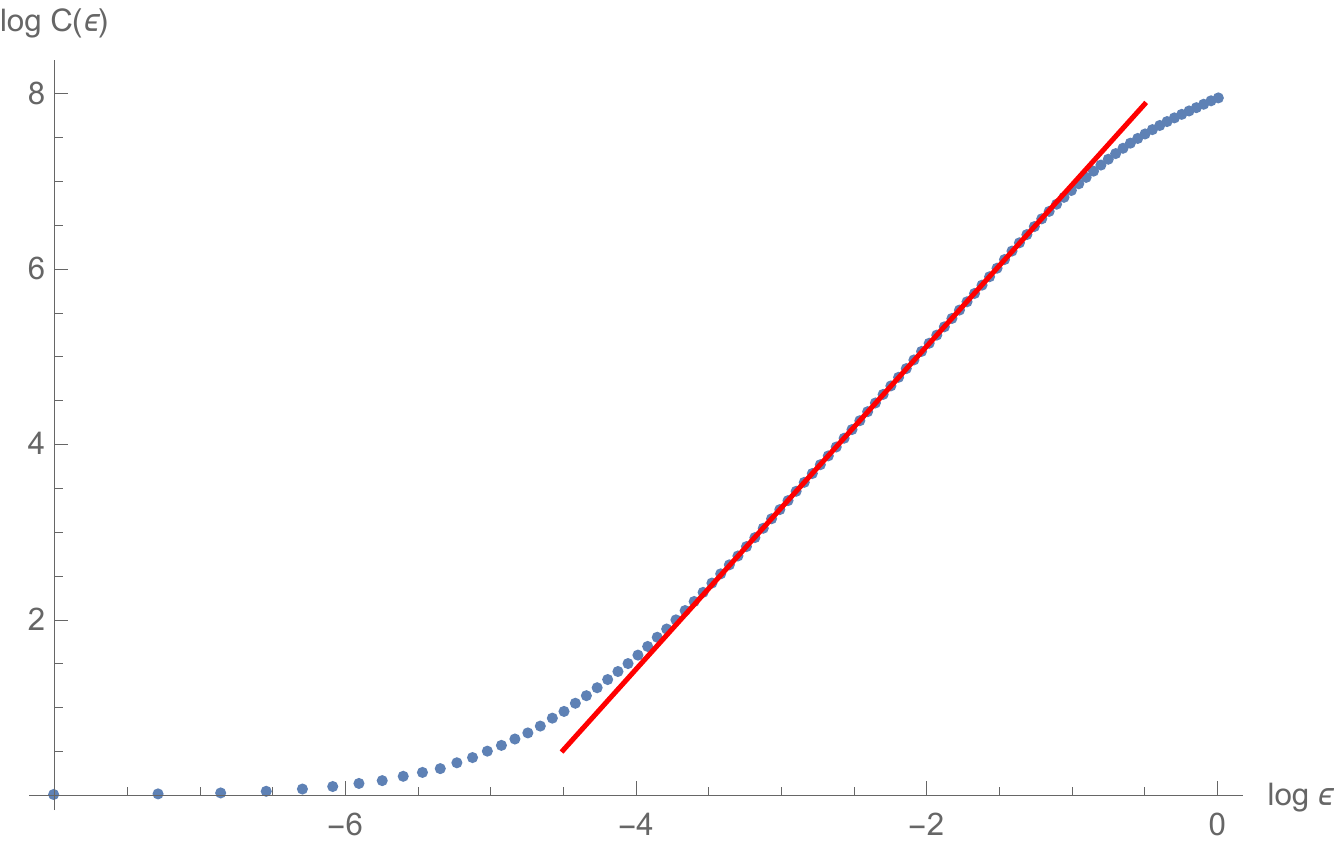}
\caption{The estimation of the correlation dimension for $r=0.75$. The slope near the inflection point is best fitted with the linear dependence given by $\log C = 8.80+1.84\log \varepsilon$.
\label{f4}
}
\end{figure}

The onset of chaos occurs at $r=r_{\infty}$. Interestingly, for $r_b<r<1$ the system returns to its stationary behaviour. This is due to a saddle-node bifurcation that gives rise to a stable fixed point $\mathbf{v}_{1}^{\ast}$. Except for two narrow regions (see next sections), for $r_{\infty}<r<r_b$ the asymptotic dynamics of the Bloch vector takes place on an attractor that is a peculiar subset of a Bloch sphere -- see Fig. \ref{f3}. This is a strange attractor whose fractal dimension can be estimated with the help of {\it the correlation dimension} \cite{GP83} in the following way. We initiate the system in a random state $\mathbf{v}_0$ and evolve it for 10 000 steps. Next, we randomly choose a point $\mathbf{w}$ on an attractor and define a ball of radius $\varepsilon$ around it. We vary $\varepsilon$ and count how many points generated by the evolution are inside this ball. We repeat this procedure for many different choices of $\mathbf{v}_0$ and $\mathbf{w}$. Finally, we calculate the average number of points $C(\varepsilon)$ inside the ball. This number should scale as \cite{GP83}
\begin{equation}
C(\varepsilon) \propto \varepsilon^d,
\end{equation}
where $d$ is the correlation dimension of the attractor. Therefore, we plot $\log C$ against $\log \varepsilon$, which should be linear for some range of $\varepsilon$, and estimate the slope -- see Fig. \ref{f4}.  We found that the correlation dimension of the strange attractor is less than two. In particular, in the case $r=0.75$ visualized in Fig. \ref{f3}, 
its value reads $d \approx 1.84$. 

\section{Lyapunov exponents and bifurcation diagram}

To describe the analyzed dynamics quantitatively we will use 
the standard notions of 
Lyapunov exponents and dynamical entropy. 
Given an initial Bloch vector $\mathbf{v}_0$ and an initial displacement $\delta \mathbf{v}_0 = \mathbf{u}_0 |\delta \mathbf{v}_0|$, Lapunov exponent reads \cite{Ott02,Vi14, PP16},
\begin{equation}
\lambda(\mathbf{v}_0,\mathbf{u}_0) = \lim_{n\rightarrow \infty} \frac{1}{n}\log \left| \mathbf{A}^{(n)}_{\mathbf{v}_0}\cdot \mathbf{u}_0 \right|,
\end{equation}
where
\begin{equation}
\mathbf{A}^{(n)}_{\mathbf{v}_0} = \mathbf{A}_{\mathbf{v}_{n-1}} \cdot  \mathbf{A}_{\mathbf{v}_{n-2}} \cdot \ldots \cdot  \mathbf{A}_{\mathbf{v}_{0}},
\end{equation}
and $\mathbf{v}_{0}, \mathbf{v}_{1}, \mathbf{v}_{2}, \ldots$ is the trajectory. Alternatively
\begin{equation}
\lambda(\mathbf{v}_0,\mathbf{u}_0) = \lim_{n\rightarrow \infty} \frac{1}{2n}\log \left(  \mathbf{u}_0^T \cdot \mathbf{H}^{(n)}_{\mathbf{v}_0}\cdot \mathbf{u}_0\right),
\end{equation}
where
\begin{equation}
\mathbf{H}^{(n)}_{\mathbf{v}_0} = \left(\mathbf{A}^{(n)}_{\mathbf{v}_0}\right)^{T}\cdot \mathbf{A}^{(n)}_{\mathbf{v}_0}.
\end{equation}
Numerical approximations give 
\begin{equation}
\lambda(\mathbf{v}_0,\mathbf{u}_0) = \frac{1}{2n}\log \left(  \mathbf{u}_0^T \cdot \mathbf{H}^{(n)}_{\mathbf{v}_0}\cdot \mathbf{u}_0 \right)
\end{equation}
for large $n$. Choosing $\mathbf{u}_0$ along the direction of the eigenvectors of $\mathbf{H}^{(n)}_{\mathbf{v}_0}$ we obtain three Lapunov exponents $\lambda_1 \geq \lambda_2 \geq \lambda_3$. 

To evaluate them numerically we used the standard procedure of
 Benettin et al. \cite{BGGS80}, described in \cite{Ott02}.

According to the Pesin theorem,
the dynamical entropy $H_{KS}$ of Kolmogorov and Sinai
is given by the sum of positive  Lapunov exponents \cite{PP16},

\begin{equation}
H_{KS} = \sum_{j=1}^J \lambda_J,
\end{equation}
where $J$ is the largest index such that $\lambda_j >0$. 
For non-chaotic systems $H_{KS} = 0$ 
while  chaotic system are defined by the condition $H_{KS} >0$.

\begin{figure}
	\includegraphics[width=8.5cm]{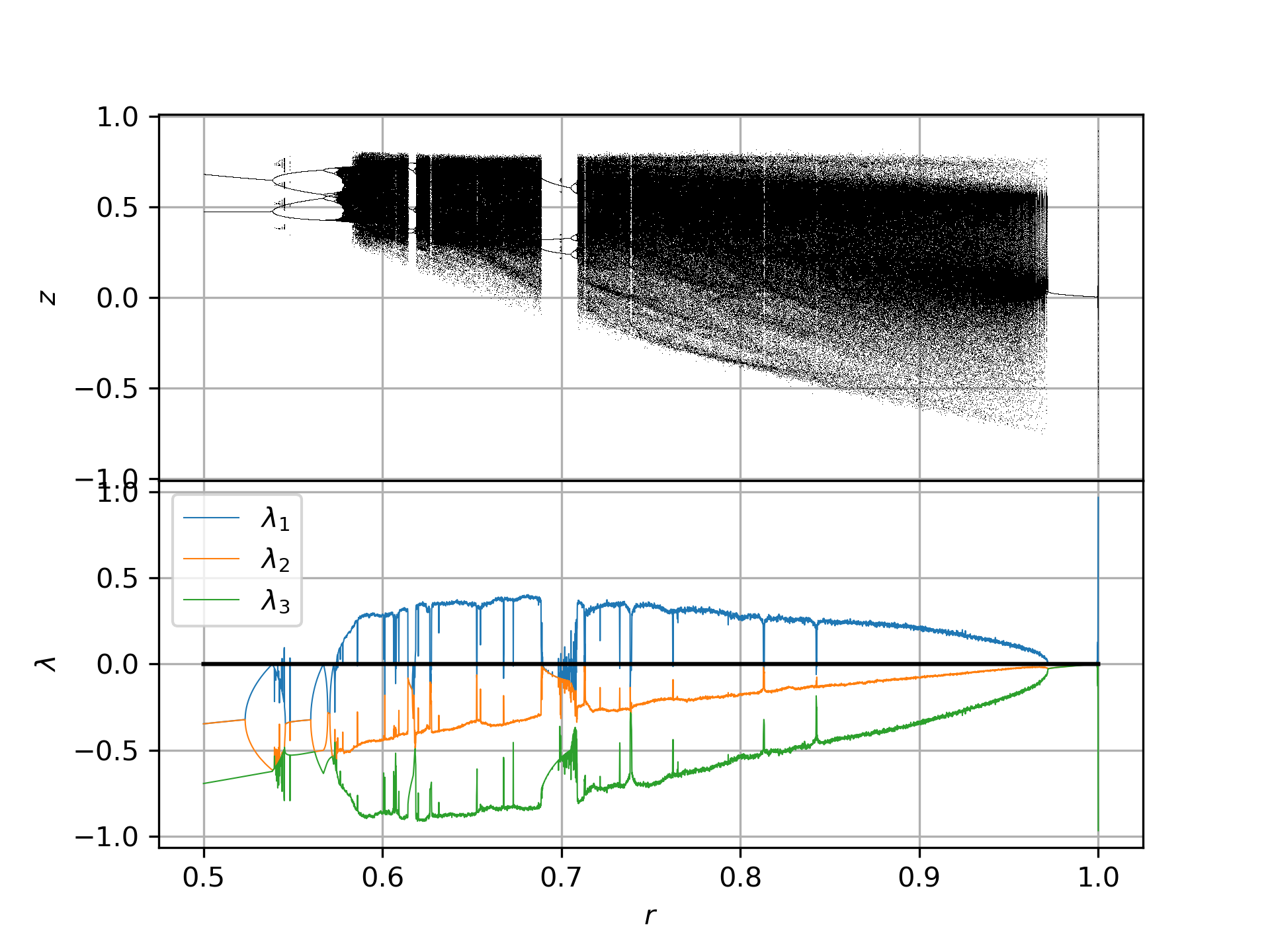}
\caption{Bifurcation diagram for the $Z$--coordinate of the Bloch vector 
(top)
to be compared with Lyapunov exponents $\lambda_j$
plotted as functions of the system parameter $r$.
The onset of chaos corresponds to positivity
of the largest exponent $\lambda_1$.}
  \label{Lyap}
\end{figure} 

\begin{figure}
     \includegraphics[width=6.5cm]{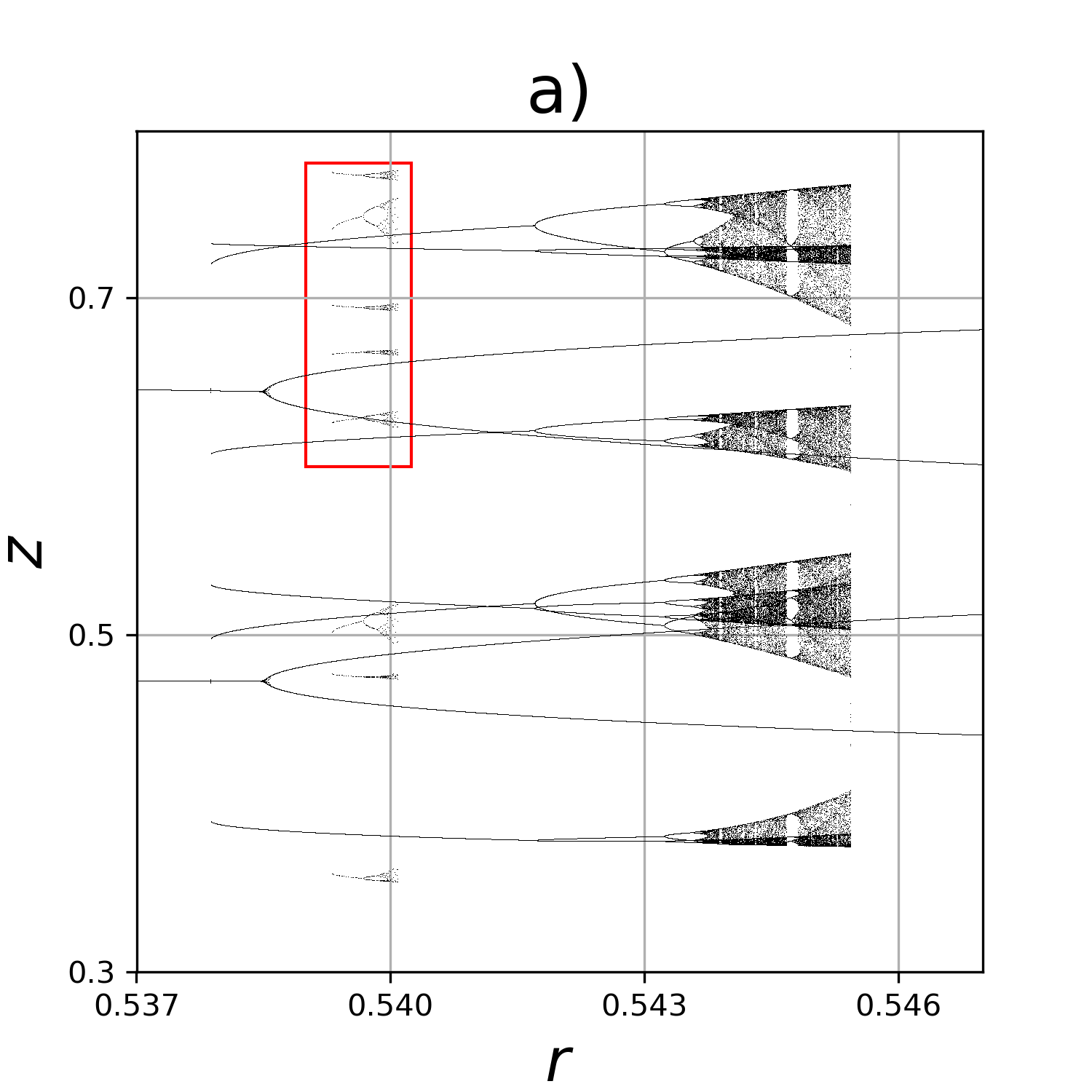}
	 \includegraphics[width=6.5cm]{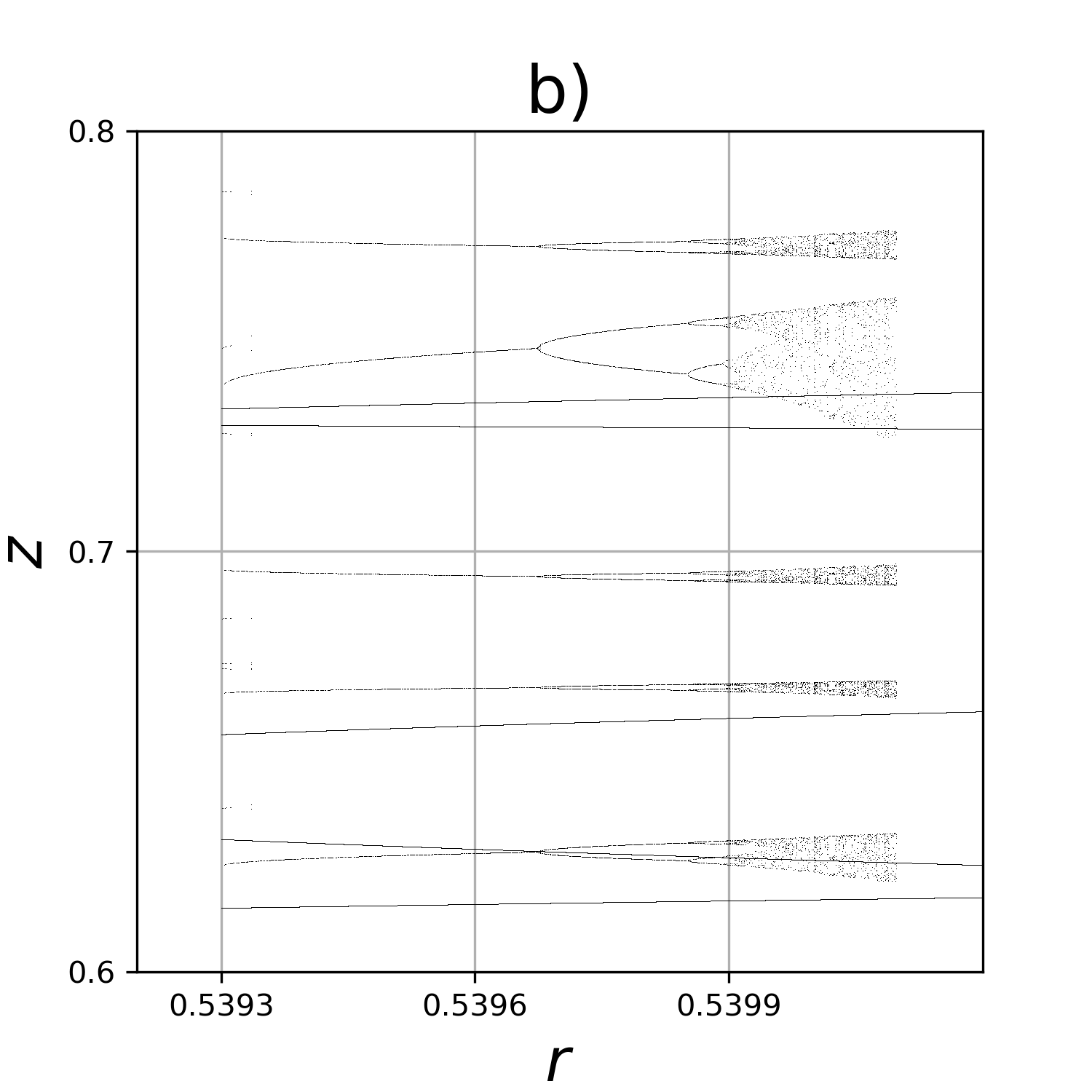}
	\caption{Magnification of Fig. \ref{cykle2}:	
	 a) secondary bifurcation diagrams occuring
	    at $r \approx 0.544$ inside the 4-cycle 
	 of the main bifurcation tree;
	  b) magnification of the rectangle from the upper panel 
	    show a ternary structure at $r \approx 0.540$.}
	  \label{self3} 
\end{figure}

Changes of the dynamics of the system as a function of the
damping parameter $r$
 is shown  in Fig. \ref{Lyap}, in which  
 the bifurcation diagram  can be compared 
 with the Layapunov exponents $\lambda_i$.
 As the second exponent $\lambda_2$ of the system
 analyzed is not positive (except very close proximity to $r=1$), the dynamical entropy $H_{KS}$,
 equal to the sum of positive exponents,
 reads in this case,  $H_{KS}=\max\{\lambda_1, 0\}$.
 Observe that the entropy is positive around $r\approx 0.7$
 at the right part of the $3$-window, 
 in which the secondary bifurcation diagram 
 leads to a small scale chaos.
 Furthermore, 
 the  system becomes (weakly) chaotic
 also at  $r\approx 0.545$
as  the secondary bifurcation scenario visible in Fig.~\ref{self3}
appear in paralell to the $4$-cycle of the main
bifurcation tree.
 
\section{Self-similarity}
\label{selfsim}
 In this section we discuss certain peculiar features of the
 bifurcation scheme of the map  (\ref{xyz+}) 
 corresponding to the quantum model studied in this work,
 which do not apear in the universal Feigenbaum bifurcation scheme,
 applicable to classical, one-dimensional  maps with a single extremum.
 In such a standard scheme one observes 
 higher order period doubling scheme 
 which occur {\sl inside} the widnows of regular motion.
 For instance, the first bifurcation inside the period-three window,
 corresponding to logistic map, leads to oscillations of period six
 and eventually leads to a small-scale chaotic dynamics
 at the right end of the window.
 Higher order diagrams can also be found 
 as entire  cascade of self-affine copies of the Feigenbaum
 bifurcation trees  can be identified -- see analysis of the 
  magnified diagrams presented in  \cite{LLBR12}.
 
 Observe, however, that  the branching pattern presented in 
 Fig. \ref{cykle2} is qualitatively different,
 as the secondary bifurcation tree localized for $r \sim  0.5455$
 appears in parallel to period-four osciallations, before the 
 main bifurcation scheme culminates in the onset of large scale
 chaotic dynamics at $r_{\infty} \approx  0.578$.
 To emphasize a selfsimilar structure of the investigated 
 bifurcation scheme we present the values of the $Z$--component of the
 Bloch vector in  magnification of the region 
  $r_{s_1} \approx 0.5378 < r < r_{s_2}  \approx 0.5455$
shown in Fig. \ref{self3}a. It is not difficult to identify  
 ternary  bifurcation structures visualized by red rectangle
 at $r\sim 0.540$.
  
\begin{figure}
	\includegraphics[width=5.5cm]{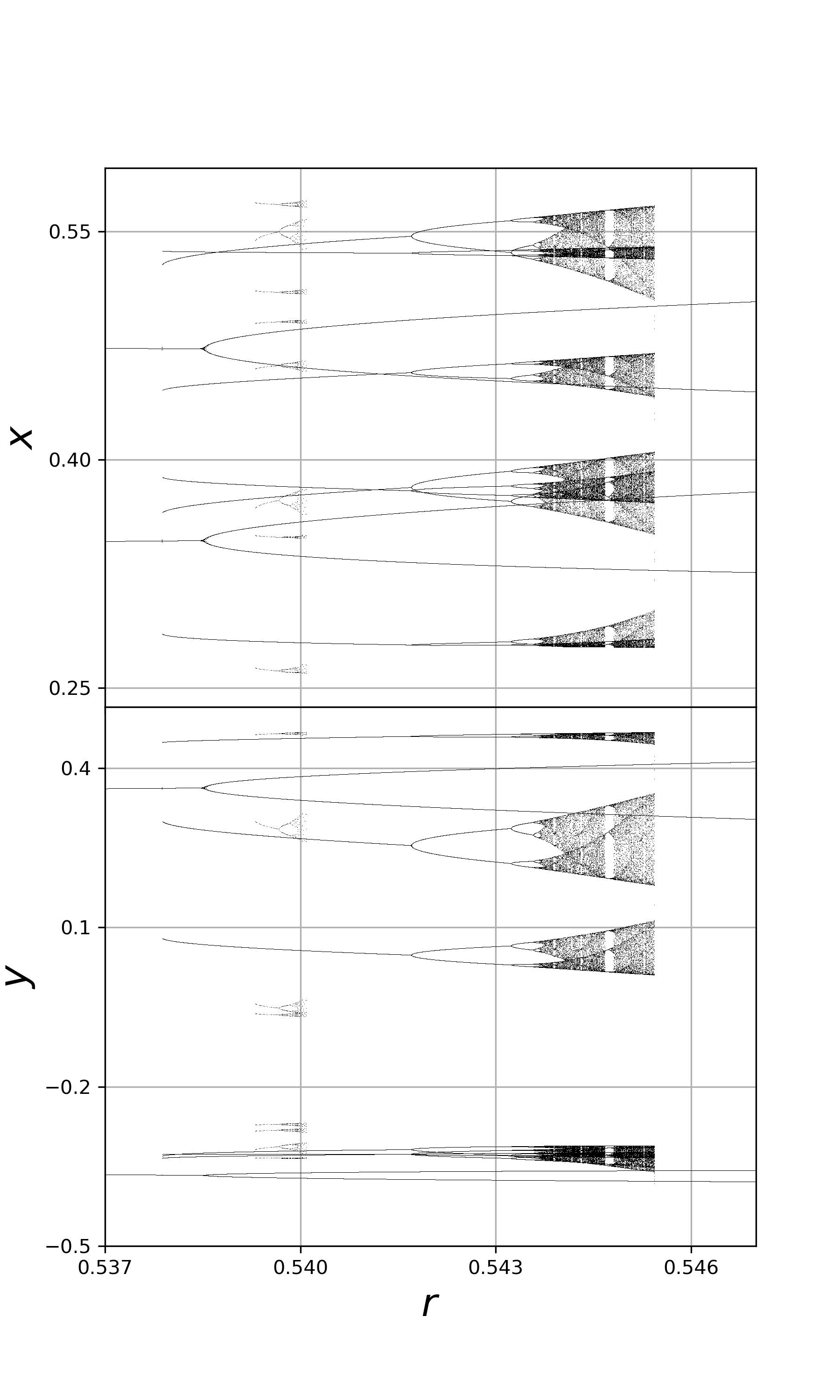}
	\caption{Secondary bifurcation diagrams for the $X$ and $Y$ 
	  coordinates of the Bloch vector visible in the same parameter range,  
	  $r \approx 0.545$, as shown in Fig. \ref{self3}a.}
	    \label{self4}
\end{figure}

Similar structures, observed for the classical, two-dimensional  H{\'e}non map \cite{ZRSM00},
can suggest that these effects are due to the fact that the analyzed map
(\ref{xyz+}) is  three dimensional.
In Fig. \ref{self4} we present behaviour of the other two components 
of the Bloch vector in the same range of the damping parameter $r$.
These results show that an analogous self-similar Feigenbaum structure
is characteristic to all three components of the Bloch vector.


\section{Concluding remarks}

In this work we investigated the system of several interacting qubits,
which realize the dynamics of the  kicked top
and undergo the damping described by two Kraus operators.
In the case the classical system is fully chaotic,
the dynamics depends exclusively on the value of the
damping parameter $r\in [0,1]$.
In the case $r=0$ the system converges to the stationary state
in a single step, while for $r=1$ (no damping)
the quantum dynamics is unitary and the corresponding 
classical dynamics is fully chaotic.
Therefore, during the parameter change  
we observe a transition from order to chaos.
Furthermore, while decreasing the damping parameter
 we identify the period doubling sequence characteristic
 to the Feigenbaum scenario, 
 originally discovered for one-dimensional dynamical systems.
  
 To the best of our knowledge, the model 
 of coupled spins subjected to the damping channel
introduced in this work
provides a first example of a quantum system, for which
the route from regular to chaotic dynamics 
according to the  universal scenario of Feigenbaum is reported.
In contrast to the standard approach,
in which the transition occurs while the non-linearity parameter 
is varied \cite{Ott02,Cvi},
in the present study the corresponding classical dynamics
is chaotic, and the period doubling takes place as 
the system parameter $r$ is increased,
so that the damping parameter $r'=1-r$ is decreased.

Interestingly, the numerical value of the 
ratio $\delta$  between consecutive values of the period-doubling
values $r_n$ of the damping parameter, 
is close to the universal Feigenbaum constant derived 
for 1-d  nonlinear transformations \cite{Fe78,Fe80}. 
It is tempting to conjecture that the observed 
transition from regular to chaotic dynamics is not
restricted to this particular model of quantum dynamics,
but it correctly describes parametric changes
of a wide class of many-body quantum systems.

As the system parameter $r$ is varied one can 
identify cycles of  oscillatory motion
and windows of periodic motion
 ordered according to the
celebrated Sharkovsky order \cite{Sha64, LY75,Sha95}.
However, we observe also self-similar structures analogous to
the entire Feigenbaum bifurcation tree,
localized in the regime of stable motion with period $4$.
Such a behvior,
earlier reported  for the two-dimensional  H{\'e}non map \cite{ZRSM00},
can be related to the fact that the investigated map
(\ref{xyz+}) is  three dimensional.

\medskip

It is a pleasure to thank Andy Chia for several inspiring discussions and helpful remarks.
Financial support by by the Foundation for Polish Science 
under the Team-Net project no. POIR.04.04.00-00-17C1/18-00, the IRAP project ICTQT Contract No. 2018/MAB/5 (cofinanced by EU via Smart Growth Operational Programme)
and by Narodowe Centrum Nauki 
under the Maestro grant number DEC-2015/18/A/ST2/00274, the Maestro grant number DEC-2019/34/A/ST2/00081, and OPUS grant number DEC-2017/27/B/ST2/02959
are gratefully acknowledged.
\newpage


\section{Appendix A}

Here we show how an effective nonlinear dynamics emerges in a multi-qudit system. Consider a single qudit in a state
\begin{equation}\label{rho}
\rho = \sum_{j,k = 1}^d \rho_{j,k}|j\rangle\langle k|
\end{equation}
and an observable
\begin{equation}
A = \sum_{j=1}^d a_j |j\rangle\langle j|. 
\end{equation}
Next, consider $N$ copies of state $\rho$, i.e., $\rho^{\otimes N}$ and a collective observable on $N$ qudits
\begin{equation}
{\mathbb A} = \sum_{n=1}^{N} A_n,
\end{equation}
where 
\begin{equation}
A_n = \openone^{\otimes (n-1)} \otimes A \otimes \openone^{\otimes (N-n)}.
\end{equation}

We are going to consider the $N$ qudit Hamiltonian
\begin{equation}
H = g {\mathbb A}^2 = g\sum_{n,m = 1}^{N} A_n A_m.
\end{equation}
This Hamiltonian is symmetric, i.e., it does not change under the permutation of qudits. Let us analyse what is the dynamics of a single qudit. The above Hamiltonian is symmetric, therefore each qudid evolves the same way, hence we can choose any qudit, say the one corresponding to $n=1$. We can rewrite the Hamiltonian as
\begin{eqnarray}
H &=& g\left(\sum_{n=2}^{N} A_n^2 + 2\sum_{n<m}^{N} A_n A_m\right) \nonumber \\
&+& g\left( A_1^2 + 2\sum_{n=2}^{N} A_1 A_n \right) = H_{env} + H_1.
\end{eqnarray} 
The part $H_1$ acts on the qudit we are interested in, whereas $H_{env}$ acts on the remaining qudits, which can be treated as an environment. Note that $H_1$ and $H_{env}$ commute (in general all the terms within these Hamiltonians commute), hence the dynamics of the system is given by
\begin{equation}
e^{-iHt} = e^{-iH_{env}t}e^{-iH_1 t},
\end{equation}
where $t$ is the time of the evolution. Therefore, the dynamics of the qudit of interest is determined by
\begin{eqnarray}
e^{-iH_1 t} &=& e^{i\frac{\chi}{2}A_1^2}e^{i\chi A_1 A_2}e^{i\chi A_1 A_3} \ldots e^{i\chi A_1 A_N}  \nonumber \\
&=& U_1 V_2 V_3 \ldots V_N,
\end{eqnarray} 
where $\chi = -2gt$.

Let us analyse the action of $V_{N}$ on the first qudit (the one we are interested in) and the N-th qudit (remember that both are in the state $\rho$ given by Eq. (\ref{rho}))
\begin{eqnarray}
& &V_N (\rho \otimes \rho) V_N^{\dagger} = \nonumber \\
& &\sum_{j,k,j',k'} e^{i\chi (a_j a_{j'}-a_k a_{k'})} \rho_{j,k} \rho_{j',k'}|j\rangle\langle k| \otimes |j'\rangle\langle k'|. 
\end{eqnarray} 
After tracing out the N-th qudit we get
\begin{eqnarray}
\rho^{(1)} &=& \text{Tr}_N \{V_N (\rho \otimes \rho) V_N^{\dagger}\} \nonumber \\ 
&=& \sum_{j,k,j'=1}^{d} p_{j'} e^{i \chi a_{j'}(a_j -a_k )} \rho_{j,k} |j\rangle\langle k| \nonumber \\
&=& \sum_{j,k=1}^{d} \gamma_{j,k} \rho_{j,k} |j\rangle\langle k|, 
\end{eqnarray} 
where $p_{j'} \equiv \rho_{j',j'}$ and
\begin{equation}
\gamma_{j,k} = \sum_{j'=1}^{d} p_{j'} e^{i \chi a_{j'}(a_j -a_k )}.
\end{equation}

Next, let us consider the subsequent action of $V_{N-1}$ on the first qudit (now in state $\rho^{(1)}$) and the $(N-1)$-th qudit (in state $\rho$)
\begin{eqnarray}
& &V_{N-1}(\rho^{(1)}\otimes \rho)V_{N-1}^{\dagger} =  \\
& & \sum_{j,k,j',k'} e^{i\chi (a_j a_{j'}-a_k a_{k'})} \gamma_{j,k} \rho_{j,k} \rho_{j',k'}|j\rangle\langle k| \otimes |j'\rangle\langle k'|. \nonumber 
\end{eqnarray} 
After tracing out the $(N-1)$-th qudit we get 
\begin{eqnarray}
\rho^{(2)}&=&\text{Tr}_{N-1} \{V_{N-1} (\rho \otimes \rho) V_{N-1}^{\dagger}\} \nonumber \\
&=& \sum_{j,k,j'=1}^{d} p_{j'} e^{i \chi a_{j'}(a_j -a_k )} \gamma_{j,k}\rho_{j,k} |j\rangle\langle k| \nonumber \\
&=& \sum_{j,k=1}^{d} \gamma_{j,k}^2 \rho_{j,k} |j\rangle\langle k|, 
\end{eqnarray}
Therefore, it is clear that after applying the sequence of operations $V_2 V_3 \ldots V_N$ the qudit of interest is in the state
\begin{equation}
\rho^{(N-1)} = \sum_{j,k=1}^{d} \gamma_{j,k}^{N-1} \rho_{j,k} |j\rangle\langle k|.
\end{equation}

Finally, let us assume that $\chi = \frac{\theta}{N-1}$, where $\theta$ is some finite constant, and that $N \rightarrow \infty$. We get
\begin{eqnarray}
& &\lim_{N \rightarrow \infty} \gamma_{j,k}^{N-1} = \lim_{N \rightarrow \infty} \left(\sum_{j'=1}^{d} p_{j'} e^{i \frac{\theta}{N-1} a_{j'}(a_j -a_k )}\right)^{N-1} \nonumber \\ 
&=& \lim_{N \rightarrow \infty} \left(\sum_{j'=1}^{d} p_{j'} \left(1+i \frac{\theta a_{j'}(a_j -a_k )}{N-1} + O(N^{-2})\right) \right)^{N-1} \nonumber \\
&=&  \lim_{N \rightarrow \infty} \left(1 + i  \frac{\theta \langle A \rangle (a_j -a_k )}{N-1}  + O(N^{-2}) \right)^{N-1} \nonumber \\
&=& e^{i\theta \langle A\rangle (a_j -a_k)}.
\end{eqnarray}
Therefore
\begin{equation}
\lim_{N \rightarrow \infty} \rho^{(N-1)} = \sum_{j,k=1}^{d} e^{i\theta \langle A\rangle (a_j -a_k)} \rho_{j,k} |j\rangle\langle k|,
\end{equation}
hence in the limit $N\rightarrow \infty$ the sequence of operations $V_2 V_3 \ldots V_N$ becomes an effective single-qudit nonlinear operation
\begin{equation}
\lim_{N \rightarrow \infty} V_2 V_3 \ldots V_N \equiv V_{nl} = e^{i\theta \langle A \rangle A}.
\end{equation}
Moreover, in the limit $N\rightarrow \infty$ we obtain
\begin{equation}
\lim_{N \rightarrow \infty} U_1 = \lim_{N \rightarrow \infty} e^{i\frac{\theta}{2(N-1)}A_1^2} = \openone,
\end{equation}
therefore we conclude that in the limit of large $N$ and weak interaction the dynamics of each single qudit is effectively governed by a nonlinear transformation
\begin{equation}
V_{nl}\rho V_{nl}^{\dagger} = e^{i\theta \langle A \rangle A} \rho e^{-i\theta \langle A \rangle A}.
\end{equation}


\end{document}